# LO-phonon assisted polariton lasing in a ZnO based microcavity

L. Orosz(1,2), F. Réveret(1,2), F. Médard(1,2), P. Disseix(1,2), J. Leymarie(1,2), M. Mihailovic(1,2), D. Solnyshkov(1,2), G. Malpuech(1,2), J. Zuniga-Pérez(3), F. Semond(3), M. Leroux(3), S. Bouchoule(4), X. Lafosse(4), M. Mexis(5,6), C. Brimont(5,6) and T. Guillet(5,6)

(1) Clermont Université, Université Blaise Pascal, LASMEA, BP 10448, F-63000 Clermont-Ferrand, France
(2) CNRS, UMR 6602, LASMEA, F-63177 Aubière, France
(3) CRHEA – CNRS, Rue Bernard Grégory, F-06560 Valbonne, France.
(4) LPN – CNRS, Route de Nozay, F-91460 Marcoussis, France.
(5) Université Montpellier 2, Laboratoire Charles Coulomb UMR 5221, F-34095 Montpellier, France
(6) CNRS, Laboratoire Charles Coulomb UMR 5221, F-34095 Montpellier, France

*Polariton relaxation mechanisms are analysed experimentally and theoretically in a ZnO-based polariton laser. A minimum lasing threshold is obtained when the energy difference between the exciton reservoir and the bottom of the lower polariton branch is resonant with the LO phonon energy. Tuning off this resonance increases the threshold, and exciton-exciton scattering processes become involved in the polariton relaxation. These observations are qualitatively reproduced by simulations based on the numerical solution of the semi-classical Boltzmann equations.*

Cavity polaritons are quasi-particles resulting from the strong coupling of excitons with photons confined in a Perot-Fabry resonator[1-3]. Polaritons are photonic particles which are however strongly interacting with each other and their surrounding media through their excitonic part. Their dispersion is remarkable, modified with respect to the one of the bare excitons and photons. This change leads to the formation of a deep and sharp energetic minimum close to the polariton ground state. These unique properties have led to many device proposals[4], the emblematic one being the so-called polariton laser[5] which does not require the achievement of the gain condition, and is therefore expected to have a very low threshold. The realisation of a polariton laser as a working device requires the condensation to occur at room temperature (RT). Wide bandgap semiconductors such as GaN and ZnO have been proposed one decade ago as appropriate candidates for RT operation[6,7], because of their large exciton binding energy and oscillator strength. RT polariton lasing has effectively been reported in bulk and multi-quantum well GaN microcavities as well as in organic microcavities[8-10]. In GaN based structures, a good understanding of the polariton relaxation mechanisms has been achieved, and the existence of optimal detuning/temperature conditions demonstrated[11]. The advantage of ZnO with respect to nitrides lies in the larger exciton binding energy and larger exciton oscillator strength leading to Rabi splitting values ($\Omega$) of the order of 120 meV[7,12] in bulk planar cavities and up to 200 meV in ZnO microwires[13]. Such a large splitting allows to keep a deep minimum of the polariton dispersion close to k=0 at zero or positive photon-exciton detuning. The polariton trap can therefore be deep (larger than 25 meV), while keeping a relatively smooth polariton dispersion which favours fast relaxation of polaritons towards their ground state through acoustic phonons[14]. The calculated optimal working temperature of such ZnO based structures should be of the order of 300K[14] instead of 170K for GaN-based structures. Another specific property of ZnO is that the LO phonon energy (72 meV at 5 K) is close to $\Omega/2$ of planar structures and therefore is expected to play a major role in the polariton relaxation mechanism. The resonance of the bottom of the lower polariton branch (LPB) with LO phonons occurs only at very negative detunings in GaN, CdTe, or GaAs based structures. In these systems, the small polariton effective mass (low density of final states) and the low excitonic fraction make the scattering process from the exciton reservoir to the ground state less effective. Strong light-matter coupling in planar ZnO microcavities (MCs) has been obtained[15-17], and recently polariton lasing reported in a negative detuning case with a low quality factor (Q~300) and polariton relaxation was found to be mainly assisted by exciton-exciton scattering[18].

In this letter, we investigate the polariton lasing effect in an hybrid ZnO microcavity at intermediate temperatures (90K and 120K) for different detunings in order to assess the main relaxation mechanisms at threshold. The lowest threshold is achieved when the energy difference between the exciton reservoir and the bottom of the LPB is resonant with the LO phonon energy, demonstrating the existence of another condensation regime, essentially phonon-assisted, with respect to reference[18]. The experimental results are interpreted within a qualitative analysis supported by theoretical calculations based on semi-classical Boltzmann equations[3,14].

The half MC (i.e. without the top mirror) has been previously studied and its potential for polariton lasing has been demonstrated[12]. The complete structure consists of a 13-pair AlN/(Al,Ga)N distributed Bragg reflector (DBR) grown on a Si(111) substrate, followed by a $5\lambda/4$ ZnO active layer and a top 10-pair SiO$_2$/HfO$_2$ DBR. The nitride DBR and the ZnO layers are elaborated by molecular beam epitaxy. Ion-beam-assisted electron-beam vacuum evaporation was used for the deposition of the SiO$_2$/HfO$_2$ mirror. It has to be noted that a SiO$_2$/HfO$_2$ DBR is used in place of the SiO$_2$/SiN mirror of the MC investigated in previous works[17,18], in order to enhance the quality factor of the cavity due to a larger refraction index mismatch. The quality factor of the cavity deduced from macro-reflectivity measurements is 150, far from the theoretical prediction (700) due to photonic disorder[19,20], but local quality factor values up to 650 have been measured on the sample. Several positions on this structure have been investigated, exhibiting different detuning values and quality factors.

Angle resolved macro-reflectivity measurements were first performed in order to highlight the strong light-matter coupling regime (SCR). A halogen lamp with a polarizer is used for the excitation and the angular resolution in detection is 1°. Figure 1a displays the evolution of the lower polariton branch (LPB) and 1st Bragg mode (B) energies as a function of the incidence angle at RT for a large detuning between the exciton and photon (δ=-65 meV). The LPB shifts towards higher energies with the increase of the incidence angle and then tends to the free exciton energy while the upper polariton branch (UPB) is diluted by the high absorption of ZnO continuum[16]. Transfer matrix simulations, reported as solid line on Fig.1a, allow to fit the experimental data in order to evaluate the Rabi splitting[21]. The removal of the band-to-band absorption in the calculations allows to reveal the UPB, and thus, Ω can be estimated to be equal to 120 meV at 300K and 125 meV at 90 K.

The angular resolved emission spectra of the cavity were then measured by using micro-photoluminescence (μ-PL) Fourier imaging. The μ-PL signal is detected through a microscope objective collecting emission angles up to 20°. The third-harmonic beam of a Ti:sapphire laser (266 nm, 130 fs pulse duration and 76 MHz repetition rate) is used for the non resonant excitation and the spot diameter on the sample is estimated to be about 2 μm. The main advantage of this configuration is the possibility to investigate the angular dependence of the emission within a small area homogeneous in term of quality factor and cavity mode wavelength[19]. Figures 1b to 1f display the evolution of the far-field emission of the cavity at 90K for a low detuning (δ=-15 meV) and for five excitation intensities. The dispersion curves of the uncoupled cavity and excitonic modes (dashed lines) together with the LPB dispersion curve (solid line) are reported on figure 1b; they are calculated from the fit of the experimental data within a coupled-oscillator model. At low excitation power, the emission is continuously distributed in momentum (or angle) along the LPB. This dispersion curve is different from that of the bare cavity mode, confirming the SCR. The occupancy of the LPB (not shown here) as a function of energy (corresponding to emission angles from 0° to 20°) reveals the absence of a bottleneck. With the increase of the excitation power, the emission from the ground state becomes dominant and a non-linear effect occurs at the threshold power ($P_{th}$ = 0.7 mJ.cm$^{-2}$ per pulse) with a strong increase of the intensity and narrowing of the emission line. A linear blue shift of the LPB as a function of the excitation intensity is observed, which is analyzed below. It is worth to mention the flat dispersion of the lasing polariton mode emission. As already reported in CdTe and GaN-based MCs[22,23], this reflects the strong localization of the condensate due to photonic disorder. From the measured angular FWHM of the polariton mode (18°) the localization radius is estimated ($r_{loc}$~ 0.4 µm). The 90 K µ-PL spectra from 0.06 $P_{th}$ to 1.2 $P_{th}$ recorded at normal incidence are reported in figure 2a. Using the linewidth corresponding to the lowest excitation power spectrum and the photonic weight deduced from the 2 coupled oscillators model, the cavity quality factor is estimated to be 600, corresponding to a photonic lifetime of 0.12 ps. At 0.06 $P_{th}$, the LPB is observed at 3299±1 meV (δ=-15meV), while the cavity mode (C) and the excitonic transition (X) are respectively at 3353 and 3368 meV (Fig. 1b). The blue shift as a function of the excitation intensity (Fig. 2a) is attributed mainly to the bleaching of the exciton oscillator strength and also to the polariton-polariton interaction[24]. Even if we are not able to quantify independently the magnitude of each effect, the exciton-exciton interaction should be negligible, according to power dependent µ-PL measurements performed on a bulk ZnO layer where no shift of the excitonic transitions is detected (not shown here). In this structure, the direct observation of the renormalization of the dispersion curve is not possible since the UPB is not detectable in ZnO layers thicker than 120 nm[16]. The following observations related to the blue shift give evidence that the system is still in SCR: (i) its evolution is linear from 0.06 $P_{th}$ to 2.12 $P_{th}$[25]; (ii) at the threshold, it is rather small (13 meV) compared to the Rabi splitting (Ω/10); (iii) even at large excitation intensity, the polaritonic emission energy is far from the cavity mode and the bulk exciton line.

A shoulder persists at low energy (3300 meV) between 0.59 $P_{th}$ and 1.12 $P_{th}$. It is probably due to the emission coming long after the pulse arrival and which is present in our time integrated measurement. Note that this increases artificially the FWHM of the LPB below threshold, where a significant blue shift is already measured. The integrated intensity and the FWHM of the main line are reported as a function of the excitation intensity (P) in figure 2b. Below threshold, the evolution of the integrated intensity is directly proportional to P indicating a polariton relaxation assisted by phonons. Indeed, at 0.88 $P_{th}$, the LPB is resonant with the first LO phonon replica (60 meV at 90 K) of the free exciton and for this detuning the excitonic part of the polariton is 0.45 which ensures an efficient scattering process. At threshold, the integrated intensity increases by 2.5 decades for a raise of the excitation intensity by only a factor 1.4. The linewidth of the LPB decreases at $P_{th}$ from 7 to 2 meV, if one consider the low density linewidth of the LPB. This attests the coherence of the polaritonic system, this last value being much lower than the estimated linewidth of the uncoupled photonic mode (5.5 meV). The increase of the LPB linewidth above the threshold is due to the achievement of the polariton condensation in different localized modes having slightly different energies and could also be related to the decoherence induced by the polariton self interactions[26,27].

Similar investigations have been carried out for a larger negative detuning point (-52 meV) at 120 K; the corresponding µ-PL spectra are displayed on figure 3a. The estimated cavity quality factor is quite similar in both experiments. The photonic and the excitonic parts of LPB are respectively 0.7 and 0.3 (θ=0°). In the "0.034 $P_{th}$" spectrum (Fig. 3a) the LPB lies at 3267.5 meV, i.e. 39.5 meV deeper than the exciton LO phonon replica (X-LO). Even at the highest intensity, the LPB is observed at 16 meV below X-LO. By comparing the two experiments (δ=-15 meV, T=90K and δ=-52 meV, T=120K), many similarities are observed: (i) at the threshold there is a strong narrowing of the emission line (down to 2 meV); (ii) the blue shift (19 meV at threshold) varies linearly with the excitation intensity below and above threshold; at high excitation power the laser line is still distinct from the uncoupled cavity mode (the energy difference is 20 meV); (iii) the shoulder is also present at the low energy side of the spectra. The relaxation mechanism here is however different from the lower-detuning case: below the threshold, the integrated intensity varies proportionally to $P^{1.8}$ (Fig. 3b). This behavior shows that the efficiency of the exciton-LO phonon relaxation mechanism is reduced with respect to the resonant case and that exciton-exciton (X-X) interactions are involved. This results in a higher polariton lasing threshold (factor 1.8).

These features are well reproduced by simulations based on the solution of semi-classical Boltzmann equations[14], presented in figure 4. These simulations show the ground state occupation versus pumping for δ=-15 meV, T=90K and δ=-52 meV, T=120K, respectively. In the low pumping range, both curves are linear, but the ground state at $\delta$=-52 meV is less occupied because of the lower efficiency of the LO-phonon assisted relaxation mechanism. When the pumping increases, the LPB population rises quadratically with the exciton density, through X-X scatterings. However, the contribution of this mechanism is only visible at more negative detuning, whereas in the case of $\delta$=-15 meV the relaxation is dominated by phonons up to the threshold. A deep polariton trap of about $9k_BT$ (T=90 K) and a resonant phonon-assisted relaxation mechanism are specific to ZnO based structures. Only these very efficient relaxation mechanisms allow the achievement of polariton lasing in structures with low Q values, 2 times smaller than in nitrides, 5-10 times than in selenides, and 50-100 times smaller than in arsenides. In the more negative detuning case the linear dependence progressively transforms in a quadratic dependence, and the stimulated threshold is about 2.5 times larger with respect to the more positive detuning case, in fair agreement with experiment. We point out that bi-exciton resonances and 2s exciton states are not considered in the simulations and as a result, the P-band[28] effect invoked in the case of ZnO exciton lasers[17] is neglected. This is consistent with μ-PL measurements performed on a bulk ZnO layer, where, in the pumping power and temperature range used, the P-band gain was absent. The comparison between theory and experiment appears to be quite satisfactory. However, one should keep in mind that simulations based on semi-classical Boltzmann equations assume a homogeneous system, whereas localization and mode competition between localized states play a central role in the definition of the states where lasing is taking place, as a result only a qualitative agreement between theory and experiment can be expected.

From the investigation of other parts of the sample we noticed that the polariton lasing is only observed for negative detunings and that no non-linear threshold has been observed between 120 and 300 K in the investigated excitation intensity range. This could be explained by the difficulty in finding a position with the required detuning and with a good quality factor because of the large photonic disorder in our sample.

The polariton lasing has been investigated in a ZnO hybrid microcavity under femtosecond excitation up to 120K. In this cavity specially designed with a $SiO_2/HfO_2$ DBR to enhance the quality factor, some photonic disorder still remains due to the nitride DBR. As a consequence, we had to investigate small areas of the sample with locally high quality factor and various detuning values. Through simulations we demonstrate the occurrence of two dominant relaxation mechanisms. When the resonance between the LPB and the LO-phonon replica of excitons is achieved, an efficient polariton relaxation toward the ground state is obtained, leading to a low threshold. A linear dependence of the integrated intensity of the emission versus the excitation intensity is then observed below threshold. Away from this resonance, exciton-exciton interactions get involved as indicated by the $P^{1.8}$ dependence of the integrated intensity. The efficiency of relaxation is reduced while the threshold value is increased. This study highlights the relative contribution of these polariton relaxation mechanisms and their importance for the threshold optimization.

The authors acknowledge J.-Y. Duboz for a critical reading of the manuscript and fruitful discussions. This work has been supported by the ANR under the ''ZOOM'' project (ANR-06-BLAN-0135) and the European Union under the ''Clermont4'' project (235114) and under the FP7 ITN "Spin-Optronics" (237252).

## Figures captions

**Figure 1.** (1a) Dispersion curves of the LPB and 1$^{st}$ Bragg mode (B) derived from angle resolved macro-reflectivity measurements performed at RT and corresponding to a large detuning value (-65 meV). The experimental data (full squares) are fitted by transfer-matrix simulations (solid line). (1b-1f) Far-field emission measured at 90K for five excitation intensities and a detuning δ = -15 meV. Dashed lines correspond to the uncoupled cavity mode (C) and the excitonic resonance (X); the solid line corresponds to the LPB calculated using the 2-coupled oscillator model.

**Figure 2.** (2a) μ-PL spectra at 90 K under normal incidence for various excitation intensities (δ=-15 meV). The inset displays the blue shift of the LPB versus power. (2b) Integrated intensity of the emission line and its linewidth reported as a function of the excitation intensity.

**Figure 3.** (3a) μ-PL spectra at 120 K under normal incidence for various excitation intensities (δ=-52 meV). The inset displays the blue shift of the LPB versus power. (3b) Integrated intensity of the emission line and its linewidth reported as a function of the excitation intensity.

**Figure 4.** Ground state occupancy as a function of pumping intensity calculated using semi-classical Boltzmann equations for a detuning of -15 meV at T=90K (black line) and -52 meV at T=120K (red line). Dashed lines indicate linear and quadratic slopes.

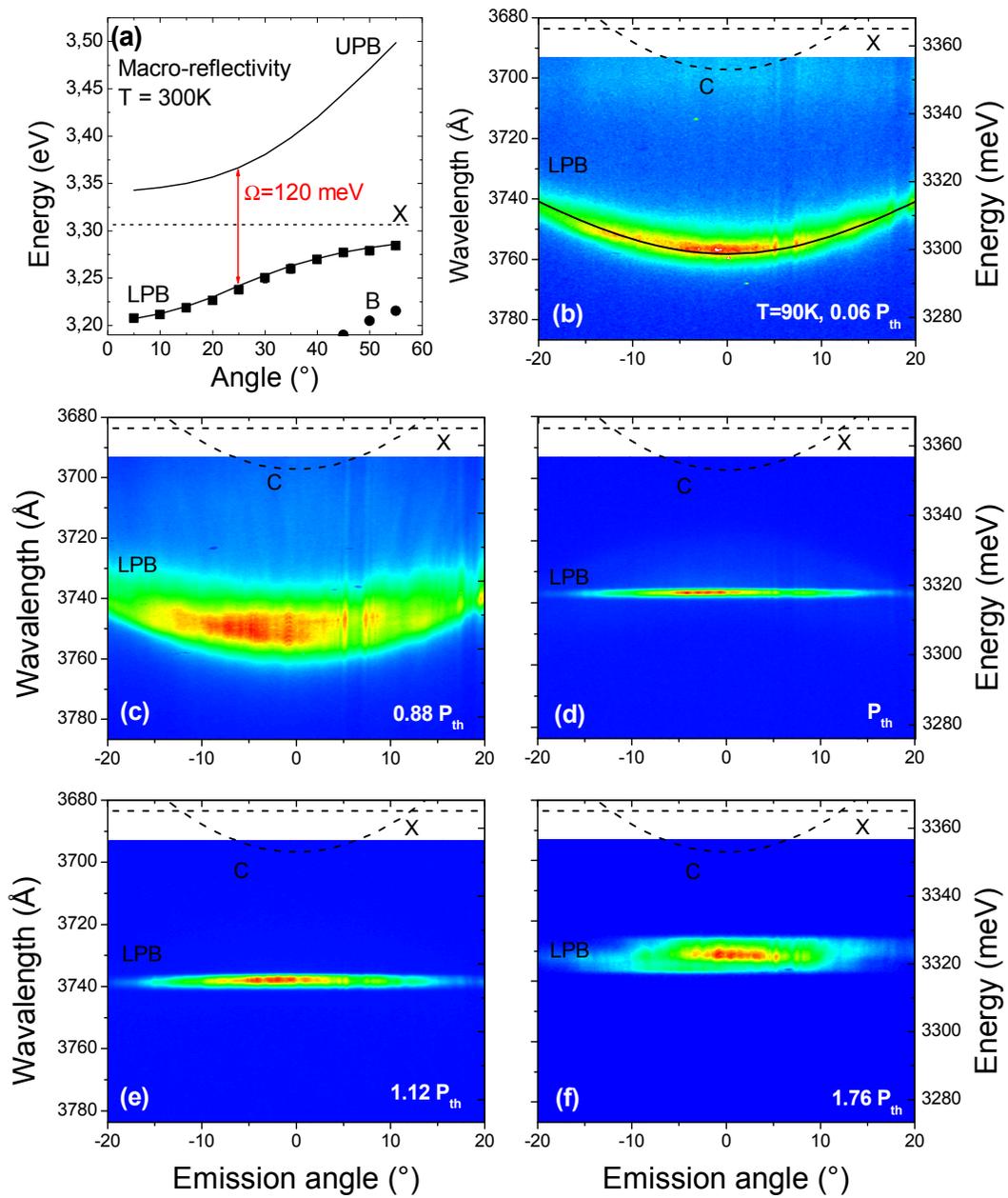

Figure 1

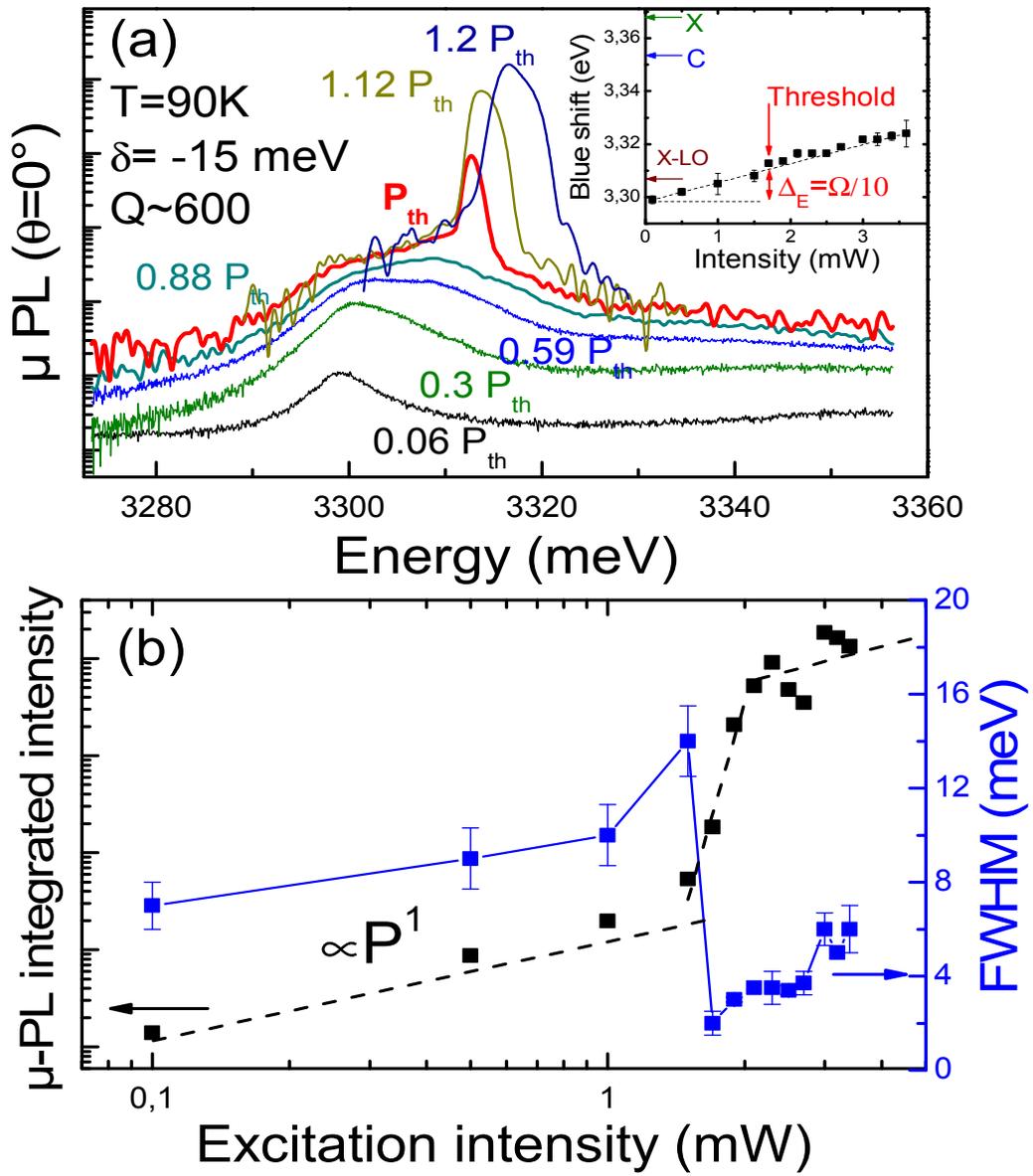

Figure 2

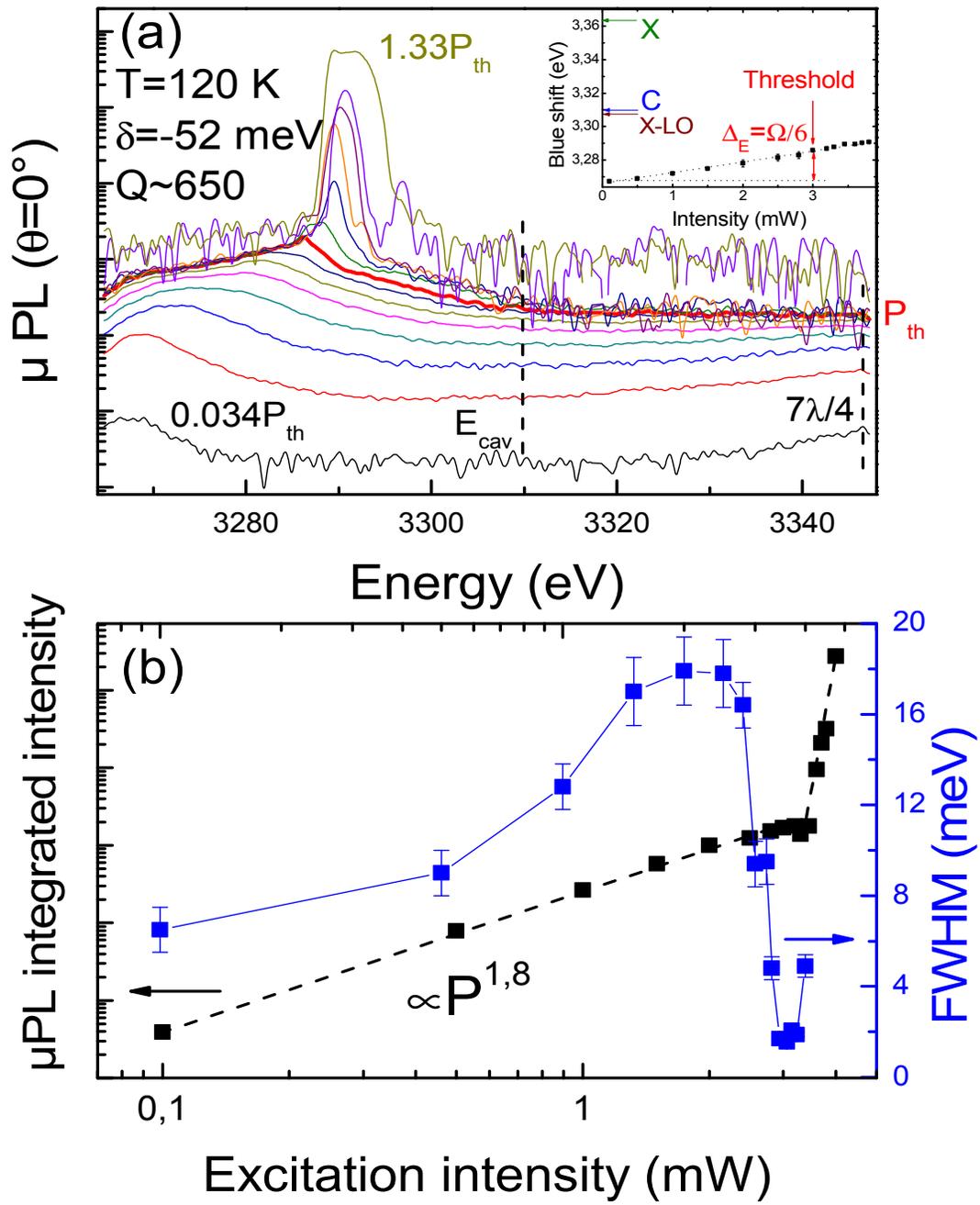

Figure 3

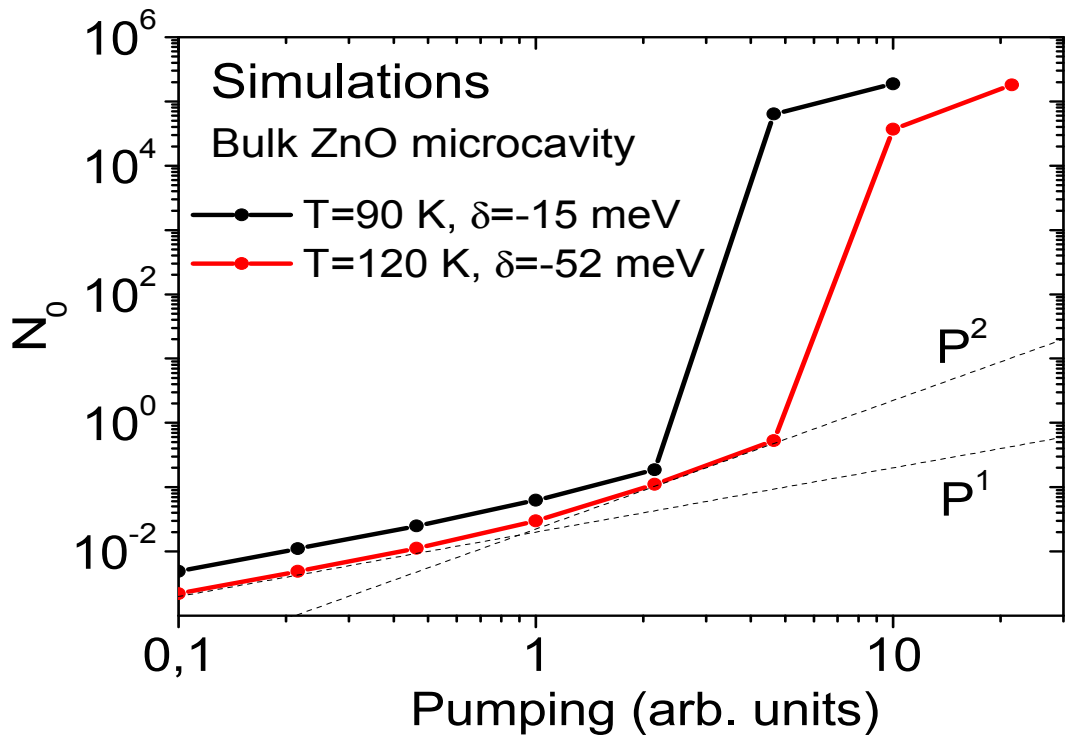

Figure 4